# Spreadsheet Engineering: A Research Framework

*Thomas A. Grossman*
*School of Business and Management, University of San Francisco,*
*San Francisco, California, USA 94117-1045*
*tagrossman@usfca.edu*

ABSTRACT

*Spreadsheet engineering adapts the lessons of software engineering to spreadsheets, providing eight principles as a framework for organizing spreadsheet programming recommendations. Spreadsheets raise issues inadequately addressed by software engineering. Spreadsheets are a powerful modeling language, allowing strategic rapid model change, and enabling exploratory modeling. Spreadsheets users learn slowly with experience because they focus on the problem domain not programming. The heterogeneity of spreadsheet users requires a taxonomy to guide recommendations. Deployment of best practices is difficult and merits research.*

1. INTRODUCTION

People have programmed computers for at least five decades. Over this time there has emerged a field called "software engineering" that considers the myriad approaches people take—and should take—when they write computer programs. This knowledge includes journal publications describing theoretical research, laboratory experiments, field observations, and recommended practices, as well as industry wisdom codified in books and computer magazines.

Since a spreadsheet is nothing more than a computer programming tool, one hopes that some of the accumulated knowledge of software engineering is relevant to spreadsheets, and [Panko 2000a] recommends we start to adapt traditional programming techniques to spreadsheets. [Rajalingham et al 2000] take a step in this direction with design recommendations and a formal hierarchical tree technique. However, we are unaware of any systematic consideration of how software engineering principles could apply to spreadsheets.

The application of software engineering principles to spreadsheets—call this "*spreadsheet engineering*"—has the potential to increase the productivity of spreadsheet programmers, decrease the frequency and severity of spreadsheet errors, enhance spreadsheet maintainability over time, and actually be implemented by spreadsheet users.

The contributions of this paper are to present the principles of spreadsheet engineering; show how spreadsheets raise engineering questions not addressed by existing software engineering results (spreadsheets as a modeling language); use software engineering ideas to explain a vexing spreadsheet research result (spreadsheet programmer experience does not correlate with accuracy); propose that researchers require an improved taxonomy of spreadsheet users; and explain why the deployment of research results is itself a research topic.

2. SOFTWARE ENGINEERING

Software engineering is a large field with decades of history, countless journal articles, numerous textbooks and professional books, and a strong presence in university teaching (sixteen semester-length software engineering courses are listed in the University of Calgary 2001-2 Calendar). There is no satisfactory definition of software engineering—in fact, [Pfleeger 2001] doesn't even try to define it. [Institute for Information Technology 2000] provides several definitions, including this gem: "software engineering is what software engineers do". For our purposes, software engineering is concerned with all aspects of software creation.

Software engineering encompasses a broad range of issues. A partial list includes planning, design, requirements specification (figuring out what the software should do), prototyping, architecture, coding, error prevention, testing and debugging (identifying, finding and fixing errors), psychology of programming, organization and management of programmers, estimating time and cost, risk management, reliability, and lifecycle management.

Research results and extensive industry experience (see for example [McConnell 1993, 1996], [Pfleeger 2001]) show conclusively that good software engineering practices increase productivity, decrease the time needed to create software, reduce the number of errors, enhance the ability to maintain and update software over time, and can be widely deployed and used given skilled managers. A central lesson is that programmers must be cognizant of the *process* by which they create software. Superficially sensible but naïve approaches have been comprehensively shown to be ineffective and costly. Good software doesn't happen spontaneously: software must be engineered.

## 3. FROM SOFTWARE ENGINEERING TO SPREADSHEET ENGINEERING

Parallel to our definition of software engineering, we provide a working I provide definition of spreadsheet engineering: "Spreadsheet engineering is concerned with all aspects of creating spreadsheets."

Decades of software engineering research and application has led to some important principles. These principles are independent of programming language, operating system, and computer hardware. Although they have not been verified on spreadsheets, we are optimistic these principles are valid for spreadsheet programs.

### 3.1. Eight Principles of Spreadsheet Engineering

**Principle 1: Best practices can have large impact**

Best practices can result in dramatically better results than naïve practices. McConnell [1996, p. 12] cites numerous studies showing productivity variations of "at least 10 to 1" for individual programmers with comparable levels of experience, and variations in the performance of entire teams "on the order of 3, 4, or 5 to 1".

**Principle 2: Lifecycle planning is important**

There is extensive research that shows planning a software project leads to dramatic improvements in cost, time, and accuracy. However, some managers are unsympathetic to programmers who spend time on planning, and some programmers don't plan because they can't resist the urge to begin coding as soon as possible [McConnell 1996 p. 23].

Whether or not the programmer realizes it, every piece of software goes through a lifecycle, starting with the conception of the software and ending the last time the software is used. A key tool is a **lifecycle model** that "establishes the order in which a project specifies, prototypes, designs, implements, reviews, tests, and performs it other activities. It establishes the criteria that you use to determine whether to proceed from one task to the next" [McConnell 1996].

The lifecycle model chosen for a project defines the master plan for the project. The lifecycle model used has significant influence over the project's success. Failure to choose a lifecycle model is itself a choice, often leading to *de facto* use of the undesirable "code-and-fix" model with unfortunate and predictable consequences.

**Principle 3: A priori requirements specification is beneficial**

It is easiest to build software when you know exactly what you want. However, specifying requirements in advance is difficult, in part because the act of programming teaches people about the problem they are trying to solve. Best practices depend on the degree to which the software requirements can be specified in advance, and make corresponding provision for learning and change during the programming process. A mismatch between the stability of the requirements and flexibility of the lifecycle model can lead to serious difficulties.

**Principle 4: Predicting future use is important**

Best practices vary with the predicted future use of the software. Software intended for one-time use by a domain expert should be constructed differently than software intended for repeat use by multiple users with heterogeneous domain knowledge. Inaccurate prediction of future use can cause increased frustration, expense, and risk.

**Principle 5: Design matters**

Designing software before coding it improves cost, time and accuracy. Effective design requires application of a set of design principles. Some design principles apply to all programming languages, such as modularity and information hiding, and visual layout showing the logical structure of the program [McConnell 1993, 1996]. Some are specific to a language, such as use of formatting and color in a spreadsheet. Design decisions depend on knowledge of what the software will do, and how it will be used.

**Principle 6: Best practices are situation-dependent**

There are no one-size-fits-all best practices. There may be more than one "smart" approach to a particular challenge. No one method is best for all situations. Any proposed best practice should clearly state the situation(s) to which it applies.

**Principle 7: Programming is a social, not an individual activity**

There is a tendency to focus on the activities of individual programmers. However, programming is a social activity, and consideration must be given to the social unit [Weinberg 1998]. Standards and practices are transmitted and supported by social interactions. In addition, there is strong evidence for the benefits of programmers working together to create code and to read, review or inspect code [McConnell 1993, 1996, Panko 1999].

**Principle 8: Deployment of best practices is difficult and consumes resources**

Adoption of proven software engineering practices varies widely within and across firms [see Maguire 1994, McConnell 1999]. A key managerial and leadership challenge is getting programmers and their supervisors to adopt best practices. Many best practices require an upfront investment of time and resources that yields net savings in the future, necessitating a disciplined approach and careful management. Motivation, teamwork, selection and training are important considerations. The social environment is an important factor.

**3.2. Spreadsheet Engineering Research Issues**

Each of the principles in the previous section raises issues for spreadsheets that merit additional research, not least on whether these principles are indeed valid for spreadsheets. It seems desirable to interpret existing spreadsheet recommendations in light of these principles.

Theoretical and practical research on spreadsheet lifecycle models has high potential to generate useful insight and provide a structure to compare existing spreadsheet programming recommendations.

Spreadsheets raise certain issues that are inadequately addressed by software engineering. We discuss these issues and the spreadsheet engineering research implications in sections 4-7.

## 4. SPREADSHEETS AS A POWERFUL MODELING LANGUAGE

Spreadsheets are a uniquely powerful business modeling language that enable modeling that would be cumbersome or impossible in other languages. In particular, spreadsheets allow rapid model changes with strategic impact, and are a uniquely effective vehicle for exploratory modeling. These issues are uncommon in software engineering.

### 4.1. Spreadsheets for Rapid Model Changes with Strategic Impact

Spreadsheets allow rapid, near instantaneous modifications to existing models. In contrast to other computer languages, it is possible to make changes to the computer code (not changes in inputs to an application, but changes to the application itself) in real time or near real time. The speed advantage for custom analyses is so large that strategic advantage can be obtained.

[Schrage 2000] provides several examples. He describes negotiations between the government and investors to sell the assets of distressed savings and loans. The government undersecretary of the treasury—himself a former top investment banker—concedes that the government was "outspreadsheeted" by the investors who "would often bring their personal computers to the negotiations…and they could model practically every disputed issue instantly."

Schrage describes an investment banking team that developed dozens of spreadsheet models overnight in response to changing acquisition offers from multiple potential buyers. "Instant analyses turned into prototypes for instant counteroffers."

These examples of rapid model change fundamentally alter the ability of firms to compete: they are of strategic importance. It is difficult to imagine any other computer language allowing such a capability. More subtly, because end-users were able to do the modeling and programming, the delays and miscommunication attendant to non-end-user programming were avoided. The spreadsheet in the hands of an end-user programmer can function as a strategic weapon.

**Implications for Spreadsheet Engineering Research**

This class of programming—high stakes, high speed coding with strategic implications—appears to be absent from the software engineering literature. There are many interesting issues regarding this sort of programming, particularly how to design a spreadsheet to ensure flexibility and reduce the likelihood of errors.

### 4.2. Spreadsheets for Exploratory Modeling

Spreadsheet users sometimes start programming with only a vague idea of what they are doing. Software engineering wisdom suggests this is poor *programming* practice. However, this can be excellent *business* practice, because exploratory modeling teaches users about their business.

When performing exploratory modeling in a spreadsheet, the spreadsheet serves as a modeling tool to structure, explore, and understand a problem; it becomes a means for expressing one's ideas. Somewhere in the process of creating the spreadsheet, users realize something they didn't know before, recognize an insight that previously eluded them, articulate key issues to colleagues, reconceptualize their problem or even figure out how to solve it. Sometimes exploratory modeling helps them define the requirements of a computer program they then construct.

Field research [Sonntag and Grossman 1999], anecdotal evidence, and my observation of business students working on case assignments in spreadsheets indicate that exploratory modeling is a powerful and important use of spreadsheets. [Schrage 2000] provides a broad discussion of the business benefits of modeling and prototyping. [Powell 1995] provides guidelines for effective exploratory modeling.

Ethnographic research supports the value of exploratory spreadsheet modeling. [Nardi 1993, p. 83] indicates that spreadsheet users "discovered needs as they went along, and the interaction with the spreadsheet package directly supported this vital process." When discussing problem-solving, she argues that users need a means to figure out what they want, and therefore "the time-consuming struggle of iteratively and incrementally developing programs is a necessary and irreplaceable component of any end user programming activity".

Exploratory modeling is consistent with the management science literature on modeling and problem-solving, which emphasizes the importance of identifying the right problem [Evans 1991 is one example] and extracting insight from models [Geoffrion 1976]. Furthermore, [Ackoff 1981] distinguishes between an expedient "resolving", an optimal "solving", and a process-change "dissolving" of a problem. Exploratory modeling—in Ackoff's words "formulating the mess"—is central to the highly desirable third approach.

Exploratory modeling is in marked contrast to the traditional view of programming. Usually, "programming" suggests the use of a computer in a purposeful manner to accomplish a *specified task*, sometimes embodied in a requirements specification. In contrast, exploratory modeling is intended to *identify the task* that is to be accomplished. The power of exploratory modeling leads to a startling conclusion: There are times when it is actually desirable for an end-user to start programming with only a vague idea of what they are doing!

**Be Careful with Artifacts**

When someone performs exploratory modeling in a spreadsheet, it looks like they are programming in a spreadsheet. They are of course in some sense programming a spreadsheet, but they are not purposefully constructing a useful computer program. They are engaging in the intellectually demanding task of modeling, with the spreadsheet serving as a vehicle for expression.

During the modeling process, exploratory modelers learn much and benefit greatly. When they are done with their exploratory modeling, they find themselves in possession of an *artifact*: a spreadsheet. This spreadsheet artifact is the residue of their inchoate modeling process. This spreadsheet artifact is intimately connected to the powerful learning the user acquired during its creation. It contains many lessons for the circumspect.

Unfortunately, this spreadsheet artifact strongly resembles a purposeful computer program. It is tempting—sorely tempting—to use this artifact as the basis for further programming. However, this artifact is not suitable for programming, because it was not designed for that purpose. If it is misused in this fashion many difficulties are certain to arise.

Although it can be desirable for a user to start exploratory modeling in a spreadsheet with only a vague idea of what they are doing, it is risky to use the resulting spreadsheet as the foundation for a purposeful program. We hypothesize that the root cause of some end-user spreadsheet problems is misuse of the artifact of exploratory spreadsheet modeling, by making it the basis for a program intended to accomplish a specified task.

**Implications for Spreadsheet Engineering Research**

Exploratory modeling in spreadsheets is important and desirable. Spreadsheet engineering practices need to accommodate it, not fight it. Nardi [1993 p. 83] recommends that "rather than attempting to provide tools that avoid the need for incrementally working problems, it will be more fruitful to develop tools that make the struggle as short, attractive and productive as possible." In particular, it may be desirable to employ a distinct spreadsheet design step after exploratory modeling but before further programming.

Some spreadsheet programming projects start as modeling projects and evolve into a mixture of purposeful programming and continued modeling. These projects will experience substantial changes during development, and spreadsheet engineering practices must accommodate these changes during development. In these situations, flexible lifecycle models have advantages over less flexible models such as the traditional sequential "waterfall model" of software development.

## 5. THE PROBLEM OF EXPERIENCE

There is an old saying that "experience is the best teacher". Common sense suggests that in most human endeavors, experience leads to higher productivity and quality, and spreadsheets ought to be no different. Alas, this is not so.

There is experimental evidence that spreadsheet users do not seem to become more effective with experience. We propose a theoretical explanation that may aid in improving spreadsheet practice.

### 5.1. A Troubling Result: Experience Does Not Matter

Spreadsheet research has generated a troubling result: increased experience using spreadsheets does *not* correlate with increased quality. [Panko 2000b, 200c] summarizes three experimental studies on the role of experience on developing and auditing spreadsheets. They show no correlation, or no significant correlation between experience and quality. In a study of debugging models with seeded errors, "Expertise increased speed but did not reduce errors. Experts caught 57% of the errors while novices caught 55%." In a study of two programming tasks, "no difference in error rates among groups…experience made little difference." In a study of a programming task, experienced subjects showed moderate improvement over inexperienced subjects, but the results were not statistically significant.

### 5.2. Substantiation from Ethnographic Research

The experiments are substantiated by ethnographic research by Nardi [1993 p. 45] who indicates spreadsheet end-user programmers do learn slowly.

> "Spreadsheets allow users to perform useful work with a small investment of time and then to go on to more advanced levels of understanding as they are ready. In our research, we found that users often *add new programming concepts to their repertoire very slowly*, all the while being very productive spreadsheet users" (italics added).

### 5.3. Amateur and Professional Programmers

This strange phenomenon is consistent with spreadsheet users being "amateur" programmers. In 1971, Gerald Weinberg wrote a classic book called "The Psychology of Computer Programming". He makes an important distinction between amateur and professional programmers [Weinberg 1998, p. 125]. He first describes how amateur programmers react to a programming challenge:

> "The amateur, being committed to the results of the particular program for his own purposes, is looking for a way to get the job done. If he runs into difficulty, all he wants is to surmount it—the manner of doing so is of little consequence."

Continuing, Weinberg describes how a professional has a different reaction:

> "Not so, however, for the professional. He may well be aware of numerous ways of circumnavigating the problem at hand…But his work does not stop there; it begins there. It begins because he must *understand* why he did not understand, in order that he may prepare himself for the programs he may someday write which will require that understanding."

Weinberg is suggesting that the task of increasing one's domain knowledge is a separate activity from increasing one's programming knowledge. The act of writing software to solve a problem is a guarantee of learning neither about the problem, nor about writing software; any learning is dependent on where the user focuses their attention.

**The Key Distinction Between Amateur and Professional**

Weinberg continues to define the key distinction between amateur and professional programmers:

> "The amateur, then, is learning about his *problem*, and any learning about programming he does may be a nice frill or may be a nasty impediment to him.
>
> "The professional, conversely, is learning about his *profession*—programming—and the problem being programmed is only one incidental step in his process of development."

These are significant differences in approach to programming. Weinberg's theory indicates a tradeoff between enhancing domain knowledge and enhancing programming knowledge.

### 5.4. The Troubling Result Explained

The experimental studies and ethnographic observation are consistent with the theoretical prediction that spreadsheet end-users, who by definition are interested in solving a problem, are unlikely to acquire professional spreadsheet programming skills with experience.

This seems to explain the troubling experimental results: experienced spreadsheet users are but amateur spreadsheet programmers. Although their experience presumably taught them much about their problems, it taught them little about spreadsheet programming.

**Research Implications**

Therefore, researchers and managers need to be cautious about interpreting the claims of self-described experienced spreadsheet programmers. We must distinguish between experience working in the spreadsheet environment, and experience studying spreadsheet programming. An instrument that measured spreadsheet programming ability would be a valuable contribution.

# 6. HETEROGENEITY OF SPREADSHEET PROGRAMMERS

Software engineering methodologies are valuable, and should be valuable when applied to spreadsheets. However, these methodologies are intended for a homogeneous group of professional computer programmers. Spreadsheet users are a heterogeneous group, and the techniques suited for one set of users will likely prove unsuitable for another set of users. To make useful recommendations we must classify spreadsheet users.

## 6.1. Existing Classifications of Spreadsheet Programmers

We are aware of two dichotomous classifications of spreadsheet users: end-user or non-end-user; and Weinberg's amateur or professional. (There are undoubtedly others, but they have not come to my attention.)

The end-user/non-end-user definition is essentially the programmer's *intention*; whether the program is written for the programmer or for someone else.

This definition is problematic because an individual is an end-user when writing a spreadsheet for his own use but is a non-end-user when writing for someone else's use. There is anecdotal evidence that end-user programmers regularly write spreadsheets for others in their workgroup. Worse, spreadsheets created for personal use are regularly transferred to other users.

The amateur/professional definition is concerned with the programmer's *attitude*: whether the programmer is focused on learning about the domain or learning about programming. Amateurs learn about programming slowly. Presumably, they are likely to make more mistakes and be less productive than professionals. The distinction between amateur and professional is not time spent programming, but time spent learning to be a better programmer.

The amateur/professional dichotomy is really a distribution. In between the extremes of the amateur who views programming as an impediment, and the professional who carefully studies every programming lesson and bug, lies a broad range of skills and interests.

## 6.2. Problems with Existing Classifications

There seems to be an inadequate vocabulary to describe important attributes of spreadsheet users. We observe a tendency to over-generalize, and to offer prescriptions that are appropriate only for a poorly-defined subset of users. This suggests that we require a richer language to classify people who use spreadsheets: we need a taxonomy of spreadsheet users.

Further evidence for the limitations of our current classification is a tendency to equate end-users and amateurs. This may have its roots in history. Before the personal computing revolution, corporate software was written solely by computer professionals because only they could access the mainframe computer. With the advent of PC's, anyone could program a computer. Computer professionals chose to use the faster, networked, more powerful mainframes while the end-users—amateurs all—were still denied access to the mainframes and used PC's. Thus, there was a time when virtually all end-users were amateurs who used PC's.

That time is past. Although today many end-users are indeed amateur programmers, there are many end-users who are skilled programmers. Some end-user programmers would undoubtedly qualify as professionals under Weinberg's definition. Some spreadsheet programmers in management consulting firms create spreadsheets for their own use, and then transfer them to clients. They know many programming techniques for their narrow range of models, but it is unlikely their techniques would be as effective applied to different classes of models.

**6.3. Towards a Taxonomy of Spreadsheet Programmers**

We need a taxonomy of spreadsheet programmers that captures a rich set of user attributes. A useful taxonomy would help us understand the range of spreadsheet programming practice, and enable us to develop, test, and deploy different prescriptions for different members of the taxonomy. It would allow us to specify the situations where existing spreadsheet recommendations should be used, and avoided.

Such a taxonomy would likely capture the differences in programmer *activities* and *training*, not merely their *intentions* and *attitudes*. It might consider the user's awareness of and ability to apply principles 2 through 5 in section 3.1. Whatever taxonomy is developed, it needs to be firmly embedded in field research that examines what spreadsheet users actually do.

**7. DEPLOYMENT OF SPREADSHEET ENGINEERING PRACTICES**

Software engineering recommendations are based on an unstated assumption: the user desires to become a better programmer. Although the software engineering literature recognizes a range of choices in a particular situation regarding which practices to use, how to deploy selected practices, the best socio-cultural environment, and actions that managers should take and eschew, the alternatives only make sense when users are conscious of the programming choices they make, and are committed to becoming more effective programmers. Software engineering practices devised for professional programmers may not be relevant to amateurs.

If spreadsheet engineering research really does have significant potential to improve the practice of spreadsheet programmers, then researchers have a duty to see it gets used. It is not sufficient merely to show in the laboratory or in a few partner organizations that improvement is possible. Improved practices need to be actively deployed to spreadsheet users.

**7.1. Spreadsheet Engineering Deployment Research Issues**

Successful deployment of a spreadsheet engineering methodology is likely to be a significant challenge. Research into how to deploy best practices is as important as research on the best practices themselves.

Social issues seem particularly important in end-user programming environments [Nardi and Miller 1990, Nardi 1993], and exploitation of existing social structures may be essential to deployment efforts.

Organizations have a *de facto* spreadsheet culture. We hypothesize that in many organizations the spreadsheet culture misperceives the difficulty of spreadsheet programming, devalues thoughtful planning and design practices, provides few incentives to learn about spreadsheet programming, and provides powerful disincentives to open discussion of important spreadsheet errors.

We need research on how end-user spreadsheet programmers can be induced to adopt better practices, particularly in decentralized organizations where the programmers are domain experts with strong personalities and political power. One approach could be to motivate users and decision makers to treat spreadsheet programming with the same level of attention and professionalism they treat the problem domain: that is, to "professionalize" the practice of spreadsheet programming.

It may be desirable to integrate research on deployment with research on best practices by explicitly developing practices that are easy to deploy. For example, tools that we can slip into users' current practice will be easier to deploy. It might be beneficial to devise "second best" practices that are likely to be voluntarily adopted.

Deployment of systems based on strict policies and enforcement of standards [such as COBIT, Butler 2000, 2001] are expensive and probably require strong centralized management. They have high potential for effective application where powerful incentives or even legal sanctions apply (there are taxonomic research questions here), but general adoption seems unlikely.

**7.2. Spreadsheet Asset Management**

Large non-spreadsheet computer programs are generally treated as valuable assets. Like any valuable asset, their purpose and use is carefully considered before purchase or construction. A group of professionals designs, builds and tests the asset. The asset is carefully deployed. People are assigned to manage the asset and maintain it over time. For spreadsheets, even sizable, valuable spreadsheets, sound asset management structures seem to be uncommon.

Spreadsheet programs can indeed be valuable assets. A spreadsheet created in 3-months full-time work by a highly-paid professional has a cost-accounting value in excess of $25,000. Anecdotal evidence suggests spreadsheet assets are under-valued and inadequately managed—just imagine treating a company car like many users treat spreadsheets of equivalent value.

The issues of *spreadsheet asset management* merit research attention. What is the value of spreadsheet assets, and how are they being managed? Are valuable assets being squandered? How can the benefits of spreadsheet engineering techniques be communicated to management and users in a way that encourages their adoption? These benefits could include more rapid acquisition of domain knowledge, higher productivity, or *measurable* benefits from reduced error rates. Return-on-investment estimates of time and money invested in spreadsheet engineering training could prove persuasive.

**7.3. The Limitations of Error Reduction as a Motivator for Change**

Reformation of spreadsheet programming practice is often justified on the basis of error reduction. This argument is problematic because of the acceptance and apparent widespread satisfaction with current practice. Although there is widespread ignorance of the extent of spreadsheet errors and the risk of naïve spreadsheet practices, all software has errors—including any alternative to spreadsheets. Tolerance of spreadsheet errors is not necessarily foolish or even irrational; it is a matter of degree and of perceived risk.

We need research into managerial perceptions toward spreadsheet risks and errors. We need research into alternative motivators for the adoption of spreadsheet engineering techniques.

**8. CONCLUSIONS**

Spreadsheet engineering is concerned with the creation of spreadsheets. Principles of software engineering provide a useful starting point for spreadsheet engineering. It seems desirable to embed existing research on spreadsheet design processes in a spreadsheet engineering framework.

This will highlight gaps in our knowledge, and identify opportunities to adapt existing software engineering knowledge to spreadsheets, reducing the time necessary to develop and deploy improved spreadsheet practices.

Spreadsheet programmers are heterogeneous; including amateurs, professionals, end-users, exploratory modelers, and rapid-change aficionados, who work in diverse cultural and managerial environments. A taxonomy of users and their attributes is necessary to develop and deploy best practices.

## 9. ACKNOWLEDGMENTS

Stephen G. Powell made valuable suggestions on a draft of this paper. Comments by three anonymous referees improved the paper and provided inspiration for future work. This paper was written while the author was at the University of Calgary and was partially supported by Canada's NSERC, grant OGP0172794. All errors are the sole responsibility of the author.

## 10. REFERENCES

Ackoff, R. (1981), "The Art and Science of Mess Management", Interfaces 11(1): 20-26.

Butler, R. (2000), "The Subversive Spreadsheet", Presentation at INFORMS Miami conference, downloadable as INFORMS_1.PDF at http://www.ucalgary.ca/mg/grossman/grossman_srig.html , accessed April 30, 2002.

Butler, R. (2001), "Applying the Cobit Control Framework to Spreadsheet Developments", European Spreadsheet Risks Interest Group Symposium Proceedings, Amsterdam, pages 7-13.

Evans, J. (1991), Creative Thinking in the Decision and Management Sciences, South-Western.

Geoffrion, A. (1976), "The Purpose of Mathematical Programming is Insight, Not Numbers", Interfaces 7(1): 81-92.

Institute for Information Technology (2000), "Software Engineering Defined!?!", http://seg.iit.nrc.ca/English/sedefn/SEdefn.html, accessed 29 April 2002.

MaGuire, S. (1994), Debugging the Development Process, Microsoft Press.

McConnell, S. (1993), Code Complete: A Practical Handbook of Software Construction, Microsoft Press.

McConnell, S. (1996), Rapid Development: Taming Wild Software Schedules, Microsoft Press.

McConnell, S. (1999), After the Goldrush: Towards a Profession of Software Engineering, Microsoft Press.

Nardi B. and J. Miller (1990), "An Ethnographic Study of Distributed Problem Solving in Spreadsheet Development", Computer Supported Cooperative Work Proceedings: 197-208. ISBN 0-89791-402-3.

Nardi, B. (1993), A Small Matter of Programming: Perspectives on End User Computing, 1993, MIT Press.

Panko, R. (1999), "Applying Code Inspection to Spreadsheet Testing", Journal of Management Information Systems, 16(2): 159-176.

Panko (2000a), "What We Know About Spreadsheet Errors", http://panko.cba.hawaii.edu/ssr/Mypapers/whatknow.htm, accessed April 30, 2002.

Panko (2000b), "Errors in Spreadsheet Auditing Experiments", http://panko.cba.hawaii.edu/ssr/auditexp.htm, accessed April 30, 2002.

Panko (2000c), "Errors During Spreadsheet Development Experiments", http://panko.cba.hawaii.edu/ssr/devexpt.htm, accessed April 30, 2002.

Pfleeger, S. (2001), Software Engineering Theory and Practice, 2$^{nd}$ Edition, Prentice Hall.

Powell, S. (1995), "The Teacher's Forum: Six Key Modeling Heuristics," Interfaces, 25(4): 114-125.

Rajalingham, K., D. Chadwick, B. Knight, D. Edwards (2000), "Quality control in spreadsheets: a software engineering-based approach to spreadsheet development", Proceedings of the 33rd Hawaii International Conference on System Sciences, Maui, Hawaii.

Schrage, M. (2000), Serious Play: How the World's Best Companies Simulate to Innovate, Harvard Business School Press.

Sonntag, C. and T. Grossman (1999), "End user modeling improves R&D management at AgrEvo Canada", Interfaces 29(5): 132-142.

Weinberg, G. (1998), The Psychology of Computer Programming, Silver Anniversary Edition, Dorset House.